# Investigation of the lead-free double perovskites Cs$_2$AgSbX$_6$ (X= Cl, Br, I) for optoelectronic and thermoelectric applications

Asim Sajjad[1], Faizan Khan[1], Shah Haidar Khan[1], Ghulam Murtaza[2],

[1]Department of Physics, University of Peshawar, Peshawar, Pakistan
[2]Department of Physics, Islamia College University, Peshawar, Pakistan

**Abstract:** Perovskite compounds have the potential to harvest solar energy as well as exploit the thermoelectric potential of a number of available materials. Here, we present the electronic, structural, thermoelectric, and optical properties of Cs$_2$AgSbX$_6$ (X = Cl, Br, I) perovskite with the help of the density functional theory (DFT). The WC-GGA approximation was used to calculate the structural parameters. All these compounds crystalize in a cubic unit cell with lattice constant increasing from 10.65 Å (Cl) to 11.14 Å (Br) to 11.86 Å (I). The mBJ-functional shows a semiconducting nature for these compounds with an indirect band gap lying at the L-X symmetry points. The optical conductivity and absorption coefficient show their peaks in the ultraviolet region, moving towards a lower energy range by inserting large size anion. The band gap of these compounds (2.08, 1.37, 0.64 eV) indicates their potential in single and multijunction solar cells. The value of refractive index at zero energy was evaluated to be 3.1, 2.2, and 1.97 for Cs$_2$AgSbCl$_6$, Cs$_2$AgSbBr$_6$ and Cs$_2$AgSbI$_6$. Effective mass of electrons is smaller than those of holes resulting in higher carrier mobility for electrons. The Seebeck coefficient, power factor, and the figure of merit were computed using the BoltzTrap code. The negative temperature coefficient of resistivity also supports the semiconductor nature of these compounds. The high electrical, small thermal conductivity, positive Seebeck coefficient, and the optimum figure of merit make these compounds suitable for thermoelectric applications.

**Keywords:** Double Perovskites, Optical properties, Thermoelectric properties.

## 1. Introduction

Energy is a basic component of society which has a key role in its development by upgrading the standards and quality of human's life. Nowadays energy crisis is a major concern of researchers globally because the energy resources are depleting continuously, which is also not sound ecologically. Fossil fuels are the primary sources of our energy demand but they are non-renewable and limited in supply. According to some estimations, they will be consumed within the next five decades [1]. The rapid increase in energy demand has motivated the scientific community to search for sustainable and cleaner alternatives. Energy from the sun and waste heat seems to be the appropriate option. Perovskite materials have the potential to harvest solar and thermal energy. Thermoelectric energy converters are also of great importance because they can utilize the waste heat being produced in power plants, factories, and heat engines [2-7].

Double perovskites are represented by the general chemical formula A$_2$BB′X$_6$ in which B and B′ represent monovalent (B$^{1+}$) and trivalent (B$^{3+}$) cations, respectively. They are named as double perovskites because their unit cell is twice that of the simple perovskites. Generally, A-site cation belongs to alkali metals (like K, Cs, Rb etc), B and B′ cations are transition metals (like Ag, Pd, Pt etc) while X-anion belongs to the halogens group.

It has been observed that the efficiency of lead-based perovskites is superior in solar cells applications [8]. Their power conversion efficiency (PCE) increased from 3.8% to 25.2% in a relatively short span of time [9, 10]. The lead-based perovskites possess the ability to be used in future photovoltaics because of their high PCE. The lead-based halide perovskites can absorb light easily in the visible range [11]. Despite their exceptional solar cell efficiency, lead is poisonous and unstable in ambient environment which prevents their further commercialization [10, 12].

The recent interest in lead-free double perovskites opens up new possibilities to find air-stable and ecologically friendly materials for solar cell applications [10]. Lead-free double perovskites materials are being investigated as safer and stable substitutes to the more successful but toxic lead-based perovskites [13].

However, figure of merit (ZT) is usually measured in order to assess and explore the thermoelectric ability of materials. Halide double perovskites are also proved to be good thermoelectric materials [11]. Halide perovskite semiconductors are preferable materials for thermoelectric applications because of their very low thermal conductivities [14].

Researchers are keen to find the best possible alternatives to lead-based perovskites. The compounds under our consideration form an important group of double perovskites. Their properties have not been fully investigated experimentally as well as theoretically. It is, therefore, imperative to perform a detailed theoretical investigation of these compounds to understand them and to judge their suitability towards optoelectronics and thermoelectric applications.

## 2. Computational Details

The structural, thermoelectric, electronic, and optical characteristics of the double perovskites $Cs_2AgSbX_6$ (X=Cl, Br, I) were investigated using the FP-LAPW [15] method within the DFT framework implemented in the Wien2K code [16]. The WC-GGA [17] and the mBJ potential [18] were used to approximate the exchange-correlation potential. Thermoelectric transport properties were calculated using the BoltzTrap code [19].

The FP-LAPW method is one of the most widely used methods to carry out electronic structure calculation. It is used to solve the Kohn-Sham's equation for many-body system. In this method, a unit cell is split into two regions i.e Muffin Tin region and interstitial region. In the Muffin Tin region, the potential is considered to be spherically symmetric, while constant in the interstitial region. Radial solutions and plane wave solutions of the Schrödinger wave equation are used in the Muffin Tin region and interstitial region, respectively. For the expansion of wave inside the spheres, the maximum value of angular momentum $l_{max} = 10$ is considered. Similarly, for expansion in the interstitial region, the plane wave cutoff value $R_{MT}.K_{max} = 7$ is chosen. The cutoff energy -6 Ry is taken to separate the core and valance states. A k-mesh with 2000 k-points in the first Brillion zone was used to perform the SCF calculations. For thermoelectric properties calculations, SCF was performed using a dense k-mesh with 5000 k-points.

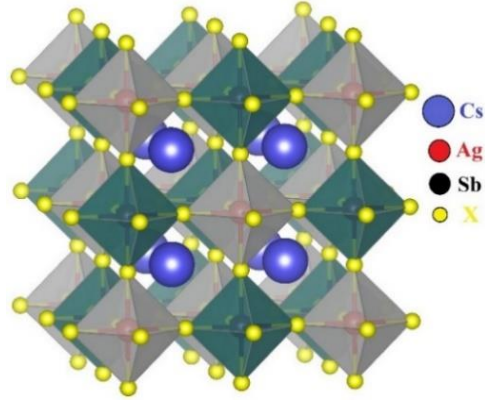

**Figure 1.** Crystal structure of $Cs_2AgSbX_6$ (X= I, Br, Cl) rendered by VESTA software.

## 3. Results and Discussion

### 3.1 Structure Properties

The structural properties of double perovskites $Cs_2AgSbX_6$ (X=Cl, Br, I) in cubic phase having space group $Fm\bar{3}m$ are investigated via volume optimization procedure. The Crystal structure is displayed in Figure 1. The GeneralizediGradientiapproximationiby Wu-Cohen (WC-GGA) was used to determine the structural parameters. The optimized $E_0$ and $V_0$ are plotted applying the Birch-Murnaghan's equation of state [20] as given below:

$$E(V) = E_0 = \frac{9V_0 B_0}{16}\left\{\left[\left(\frac{V_0}{V}\right)^{2/3} - 1\right]^3 - {B'}_0 + \left[\left(\frac{V_0}{V}\right)^{2/3} - 1\right]^2\left[6 - 4\left(\frac{V_0}{V}\right)^{2/3}\right]\right\}$$

The optimized $E_0$ vs $V_0$ plots are displayed in Figure 2 (a-c). The minimum points of these parabolic plots indicate the ground state of the compounds. The optimized structural parameters such as lattice constant $a_0 (Å)$, bulk modulus B(GPa) pressure derivative (first order) of bulk modulus ($B^p$), ground state unit cell volume ($V_0$) and energy ($E_0$) are listed in Table 1.

The values of $a_0$ increases as we change the cation from Cl to Br to I, as shown in Figure 4.2 (d) due to increase in their atomic size. The values of lattice constant agree with the experimental data listed in Table 1. An increase can be seen in the unit cell volume when we go down the group from Cl to I because of the addition of more energy levels. Amongst the three compounds, $Cs_2AgSbI_6$ has the smallest bulk modulus. Therefore, $Cs_2AgSbI_6$ is more compressible and less hard as compared to the other.

### 3.2 Electronic Properties

To investigate the electronic properties of $Cs_2AgSbX_6$ (X=Cl, Br, I), we determined the band structure (BS) and density of states (DOS) using the mBJ-functional. All the three compounds show an indirect bandgap as shown in Figure 3. The computed band gaps of $Cs_2AgSbX_6$ are listed in Table 2 2.08 eV (Cl), 1.37 eV (Br), and 0.64 eV (I), respectively, showing semiconductor nature of these materials. The substitution of X-site anion sequentially by Cl, Br, and I, reduces the band gap which can be corelated with the atomic size of these anions. Being large in atomic size, iodine contains large number of nucleons as compared to Br and

Cl. The replacement of large size anion at X-site reduces the electrostatic force on outer electrons. As a result of weak electrostatic force, the bonding energy and the separation between conduction and valance bands also decrease which reduces the band gap.

In order to determine the contribution of various electronic states, total density of states (TDOS) and partial density of states (PDOS) has been calculated and shown in Figure 4 and Figure 5. The valance band (VB) is mainly contributed by the Silver (Ag) and the corresponding halogen atoms (Cl, Br, I), whereas antimony (Sb) and the corresponding halogen atoms contributes more to the conduction band (CB). It can also be seen that at higher energy (above 5 eV), cesium (Cs) atom along with the corresponding halogen atoms contribute minimally to the CB. The lower part of the VB at energy -9 eV to -8 eV has large contribution from Cs and a little contribution from Sb atom as obvious from the graph.

The PDOS plots show that the lower VB (-8 to -9 eV) of all the three double perovskites is mostly composed of the Cs-$p$ and Sb-$s$ orbitals while the higher VB has major contribution from the Ag-$d$ and halogen $p$ orbitals. The lower part of the CB is majorly comprised of Sb-$p$ and halogen $p$ orbitals. The higher CB at the energy range 5-8 eV is composed of the Cs-$d$ orbital while the energy range 13-16 eV is mainly made up of Cs-$f$ orbitals.

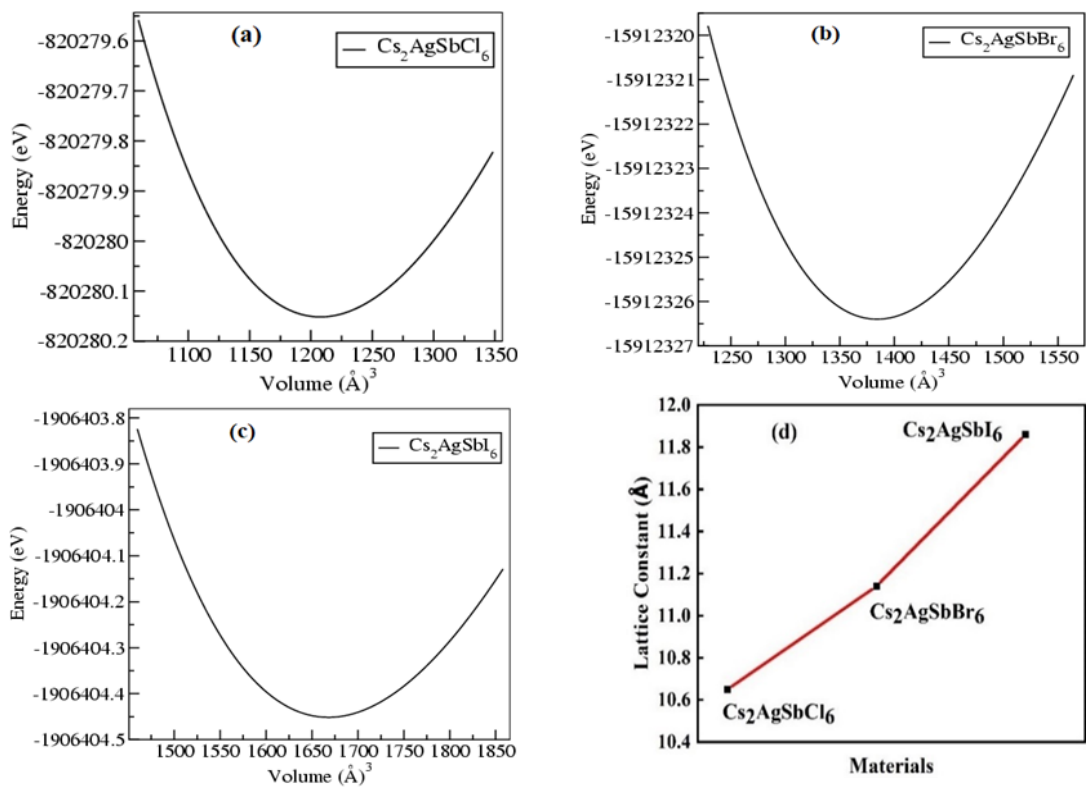

**Figure 2.** The volume optimization graphs of (a) $Cs_2AgSbCl_6$ (b) $Cs_2AgSbBr_6$ (c) $Cs_2AgSbI_6$. (d) The variation of the lattice constant with halogens (Cl, Br, I).

**Table 1.** The calculated optimized structural parameters of $Cs_2AgSbX_6$ (X=Cl, Br, I).

| Compounds | $a_0$(Å) | | $V_0$(Å$^3$) | B (GPa) | $B^P$(BP) | $E_0$(eV) |
|---|---|---|---|---|---|---|
| | This Work | Experimental | | | | |
| $Cs_2AgSbCl_6$ | 10.65 | 10.69*, 10.66** | 1207.9 | 32.09 | 5.000 | -820280.15 |
| $Cs_2AgSbBr_6$ | 11.14 | 11.15[b], 11.21[b] 11.18[a] | 1382.4 | 28.24 | 5.000 | -159123264 |
| $Cs_2AgSbI_6$ | 11.86 | | 1668.2 | 23.68 | 5.000 | -1906404.45 |

*[21], **[8], [a][22], [b][23]

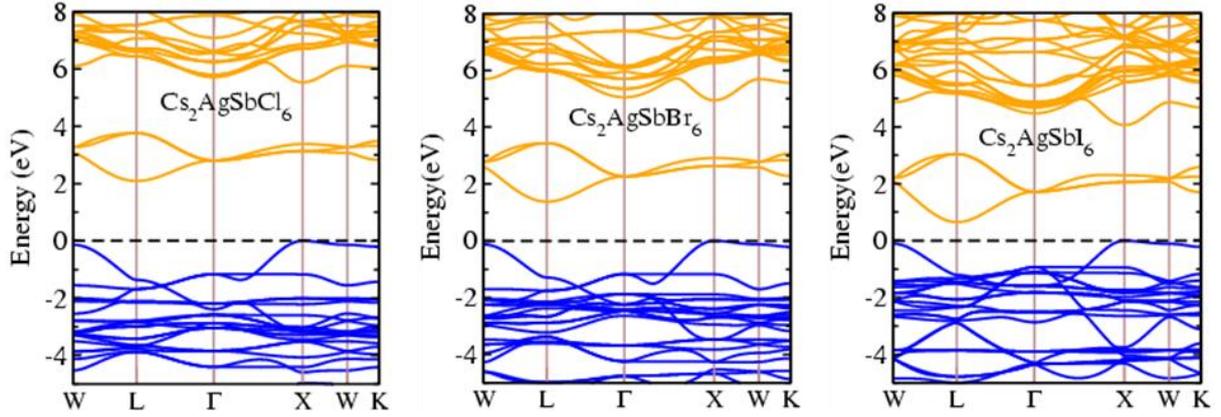

**Figure 3.** The calculated band structure of double perovskites $Cs_2AgSbX_6$ (X=Cl, Br, I).

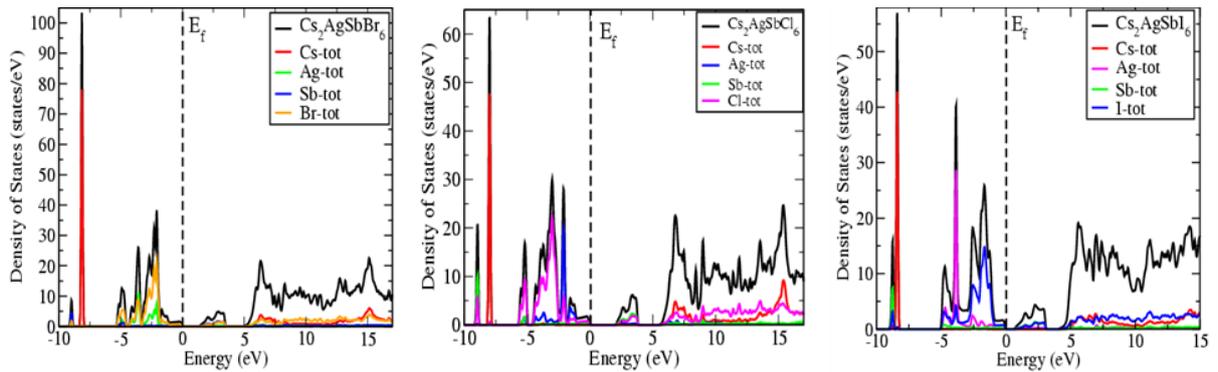

**Figure 4.** Total density of states of double perovskites $Cs_2AgSbX_6$ (X=Cl, Br, I).

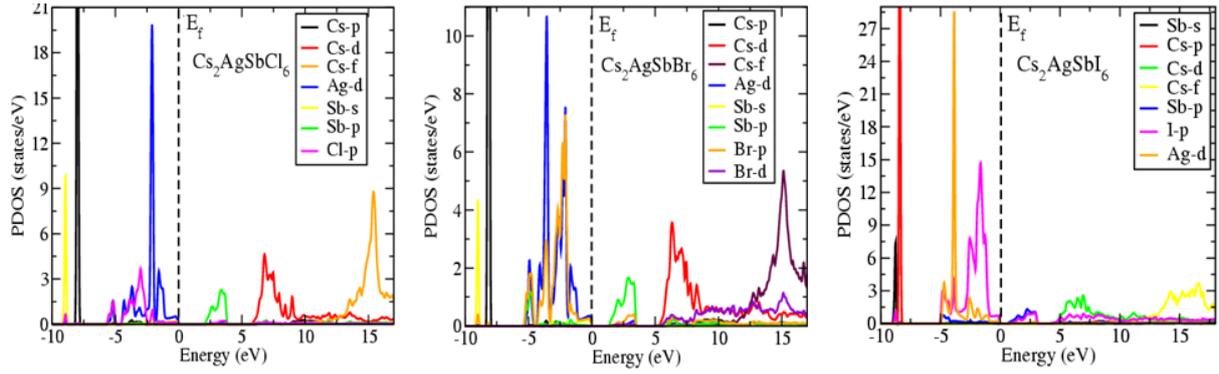

**Figure 5.** Partial density of states of double perovskites $Cs_2AgSbX_6$ (X=Cl, Br, I).

**Table 2.** Calculated bang gaps of double perovskites $Cs_2AgSbX_6$ (X=Cl, Br, I).

| Compounds | $E_g$ (eV) | | |
|---|---|---|---|
| | This work | Other work (DFT) | Experimental |
| $Cs_2AgSbCl_6$ | 2.08 | 2.22[a] | 2.54[b], 2.7[c], (2.61-2.24)[d], 2,60[e] |
| $Cs_2AgSbBr_6$ | 1.37 | 1.46[f] | 1.64[f], 1.93[g] |
| $Cs_2AgSbI_6$ | 0.64 | -- | -- |

[24][a], [25][b], [8][c], [10][d], [21][e], [23][f], [22][g]

## 3.3 Optical Properties

The interaction of light with matter can be used to analyze optical performance of materials. We have inspected optical parameters of $Cs_2AgSbX_6$ (X=Cl, Br, I) by calculating optical conductivity $\sigma(\omega)$, absorption coefficient $\alpha(\omega)$, refractive index $n(\omega)$, and effective masses. Optical properties are important for its application in optoelectronics and solar cell applications. in order to check their suitability in optoelectronic.

The optical conductivity $\sigma(\omega)$, measured in ohm per centimeter, of $Cs_2AgSbX_6$ (X=Cl, Br, I) is displayed in Figure 6 (a). The optical conductivity starts at 2.7ieV, 2.1ieV and 1.5ieV for $Cs_2AgSbCl_6$, $Cs_2AgSbBr_6$ and $Cs_2AgSbI_6$, respectively. The figure reveals that the optical conductivity increases initially and reaches a certain maximum value for each compound and then decreases again along the high energy range. The maximum value of the optical conductivity of $Cs_2AgSbCl_6$ is 4401 at 12 eV, $Cs_2AgSbBr_6$ is 4440 at 11.1 eV and $Cs_2AgSbI_6$ is 5129 at 7 eV respectively. The maximum value of optical conductivity shifts towards lower energy when anion of large size is placed. However, optical conductivity increases when we move from small sized anion Cl to large sized anion I. The broad range of $\sigma$ for all the compounds lie in a wide energy span of 6 eV to 12 eV, showing their potential to use in optoelectronics in ultraviolet range.

The absorption coefficient $\alpha(\omega)$ provides important information about a material's light harvesting ability and can affect the efficiency of solar cell. Figure 6 (b) shows the absorption coefficient α(ω) of the considered double perovskites. The absorption starts at threshold energy of 2.4 eV (Cl), 1.95 eV (Br) and 1.25 eV (I), respectively listed in Table 3, indicating optical band gaps of these materials which are in agreement with the experimental band gaps listed in Table 2. No absorption can be seen below the threshold energy which shows the materials transparency in this energy range. The optical absorption in all the three compounds starts in the visible spectrum of light and extends to the ultraviolet spectrum. The threshold values shift towards lower energy when Cl is replaced with Br and I. Among the three compounds, $Cs_2AgSbI_6$ has a peak in the visible region at about 3.05 eV.

The speed of light reduces upon entering into the material. The transparency of a material can also be evaluated in terms of refractive index n(ω) as depicted in Figure 6 (c). The zero-energy value of refractive index n(0) of $Cs_2AgSbX_6$ are 1.9 (Cl), 2.2 (Br), and 2.6 (I). Initially, n(ω) gradually increases and reaches peak value of 2.7 (Cl), 3.0 (Br) 3.5 (I). The peak values shift towards lower energy range with the replacement of Cl with Br and I. Beyond the peak values, variations occur which depicts that light enter into the material. Moreover, the variations indicate the lower transparency of the compounds in high energy region.

The incident electromagnetic radiation generated the carriers whose effective mass was estimated from the curvature of the band structure. The effective masses of electrons and holes ($m_e^*, m_h^*$) as represented in Table 2, were calculated from the curvature of CBM and VBM, respectively. It is evident from the table that effective mass of electrons is smaller than those of holes resulting in higher carrier mobility for electrons. Furthermore, it can also be seen that the electrons as well as holes effective masses decrease when we vary the halogen anion in the fashion I>Br>Cl. The larger effective mass of holes shows p-type semiconductor nature of the compounds.

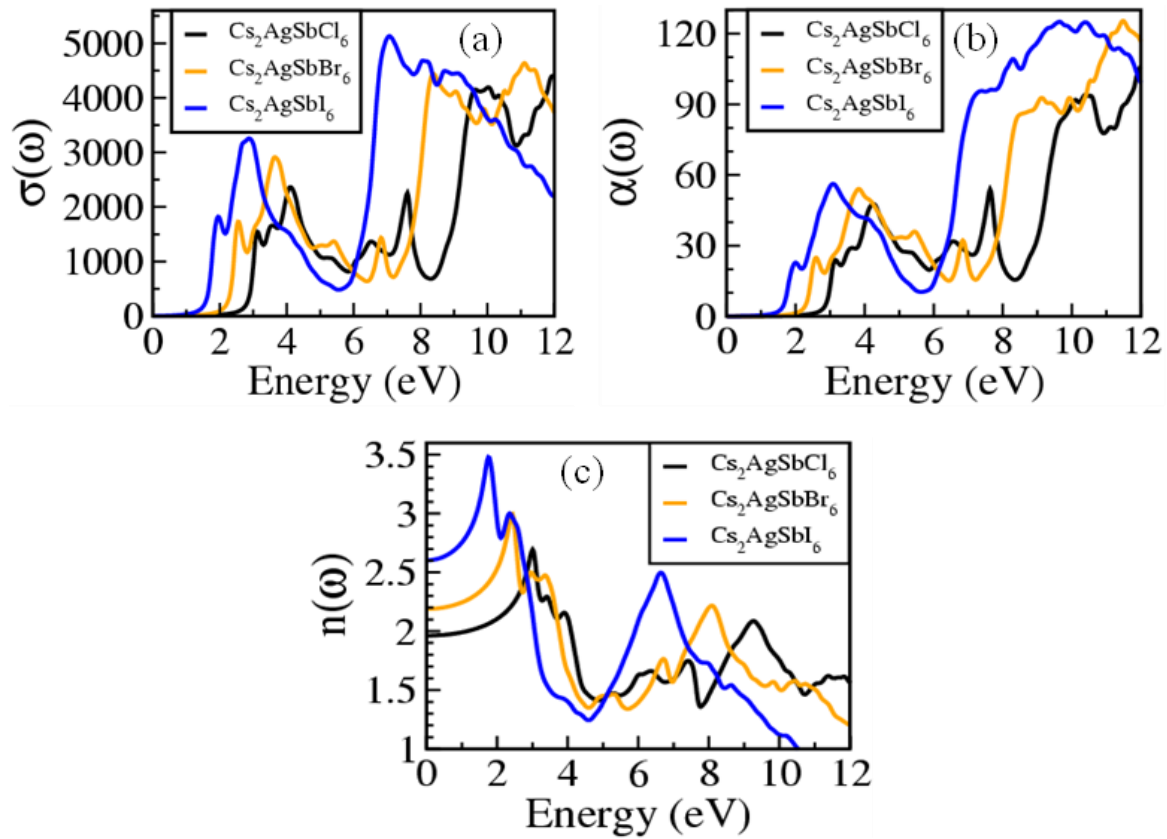

**Figure 6.** Variations of (a) Optical conductivity (b) Absorption coefficient and (c) Refractive index with increasing photon energy.

**Table 3.** The calculated optical band gaps of double perovskites $Cs_2AgSbX_6$ (X=Cl. Br, I).

| Compounds | Optical bandgap (eV) |
|---|---|
| $Cs_2AgSbCl_6$ | 2.47 |
| $Cs_2AgSbBr_6$ | 1.97 |
| $Cs_2AgSbI_6$ | 1.33 |

**Table 4.** The calculated effective mass of double perovskites $Cs_2AgSbX_6$ (X=Cl. Br, I).

| Compounds | Electron effective mass $m_e^*$ ($m_0$) | Hole effective mass $m_h^*$ ($m_0$) |
|---|---|---|
| $Cs_2AgSbCl_6$ | 0.348 | 0.364 |
| $Cs_2AgSbBr_6$ | 0.250 | 0.265 |
| $Cs_2AgSbI_6$ | 0.195 | 0.252 |

## 3.4 Thermoelectric Properties

We have investigated the thermoelectric properties of $Cs_2AgSbX_6$ (X=Cl, Br, I) double perovskites using the BoltzTrap code. Figure of merit (ZT) is calculated to evaluate the thermoelectric performance of the compounds.

The electrical conductivity σ $(Ω.m.s)^{-1}$ against temperature T (100-800 K) of the double perovskite is displayed in Figure 7 (a). The calculated results indicate a linear increase in σ with increasing temperature. We can also say that electrical resistivity decreases with rise in temperature signifying negative temperature coefficient of resistivity, which confirms the semiconductor nature of the compounds. It also suggests that rise in temperature increases the intrinsic carrier concentration. The σ value at 100 K decreases from $8.91 \times 10^{17} (Ω.m.s)^{-1}$ to $7.15 \times 10^{17} (Ω.m.s)^{-1}$ to $5.36 \times 10^{17} (Ω.m.s)^{-1}$ while moving from Cl to Br to I, seemingly because of the larger sizes of the anions. This decrease can also be attributed to the inverse relation between unit cell volume and electrical conductivity [3].

The temperature dependent thermal conductivity κ (W/mKs) is plotted in Figure 7 (b). It is evident from the graph that κ increases with temperature. There is negligible difference in κ for all the compounds up to 400 K. Beyond this temperature, the difference becomes prominent and the conductivity increases linearly. At higher temperature, $Cs_2AgSbI_6$ has the lowest conductivity among the three compounds, indicating its potential to use in thermoelectric applications as compared to the other two.

The calculated pawer factor (PF) as a function of temperature is plotted in Figure 7 (c). The PF for all these double perovskites increases with increase in temperature. Their room temperature (300 K) values decrease negligibly from $1.27 \times 10^{11} W/mK^2s$ to $1.22 \times 10^{11} W/mK^2s$ while moving from Cl to I., An appropriate change in PF occurs above 400 K and reaches its maximum value of $4.18 \times 10^{11} W/mK^2s$, $4.0 \times 10^{11} W/mK^2s$ and $3.66 \times 10^{11} W/mK^2s$ at 800 K for $Cs_2AgSbCl_6$, $Cs_2AgSbBr_6$ and $Cs_2AgSbI_6$, respectively.

The ratio of the voltage produced (del V) to the temperature gradient (del T) is known as the Seebeck coefficient (S). S plotted against temperature is shown in Figure 7 (d). Our results indicate that the value of S rises with the rise of temperature and vice versa. The value of S increases and reach to its maximum value from 0.16 mV/K to 0.22 mV/K for $Cs_2AgSbCl_6$, 0.18 mV/K to 0.23 mV/K for $Cs_2AgSbBr_6$, and 0.20 mV/K to 0.24 mV/K for $Cs_2AgSbI_6$. The maximum value for $Cs_2AgSbCl_6$, $Cs_2AgSbBr_6$ and $Cs_2AgSbI_6$ occurs at 400 K, 350 K and 300 K. In the higher temperature range, its value starts decreasing with increasing temperature because of the higher carrier concentration. There is an inverse relation of S with carrier concentration. Positive values of S are observed over the whole range of temperature (100K-800 K), which show their p-type semiconductor nature [4]. An increase in S can be seen with the replacement of X-site anion with Cl, Br and I, due to the fact that bulkier ions would greatly decrease the movement of carriers during thermal agitation and hence would enhance the potential difference.

Figure of merit (ZT) as a function of temperature is illustrated in Figure 7 (e). ZT increases with temperature for all the three compounds. The ZT value at 100 K is 0.57, 0.62 and 0.66 for $Cs_2AgSbCl_6$, $Cs_2AgSbBr_6$ and $Cs_2AgSbI_6$, respectively. The double perovskite $Cs_2AgSbI_6$ has higher ZT value as compared with the others. While these values of ZT increase and reach to their maximum of 0.77 at 500 K for $Cs_2AgSbI_6$, 0.76 at 700 K for $Cs_2AgSbBr_6$ and 0.76 at 750 K for $Cs_2AgSbCl_6$, respectively. These double perovskites are good candidates to use in thermoelectric devices, based on their ZT values obtained in our calculations.

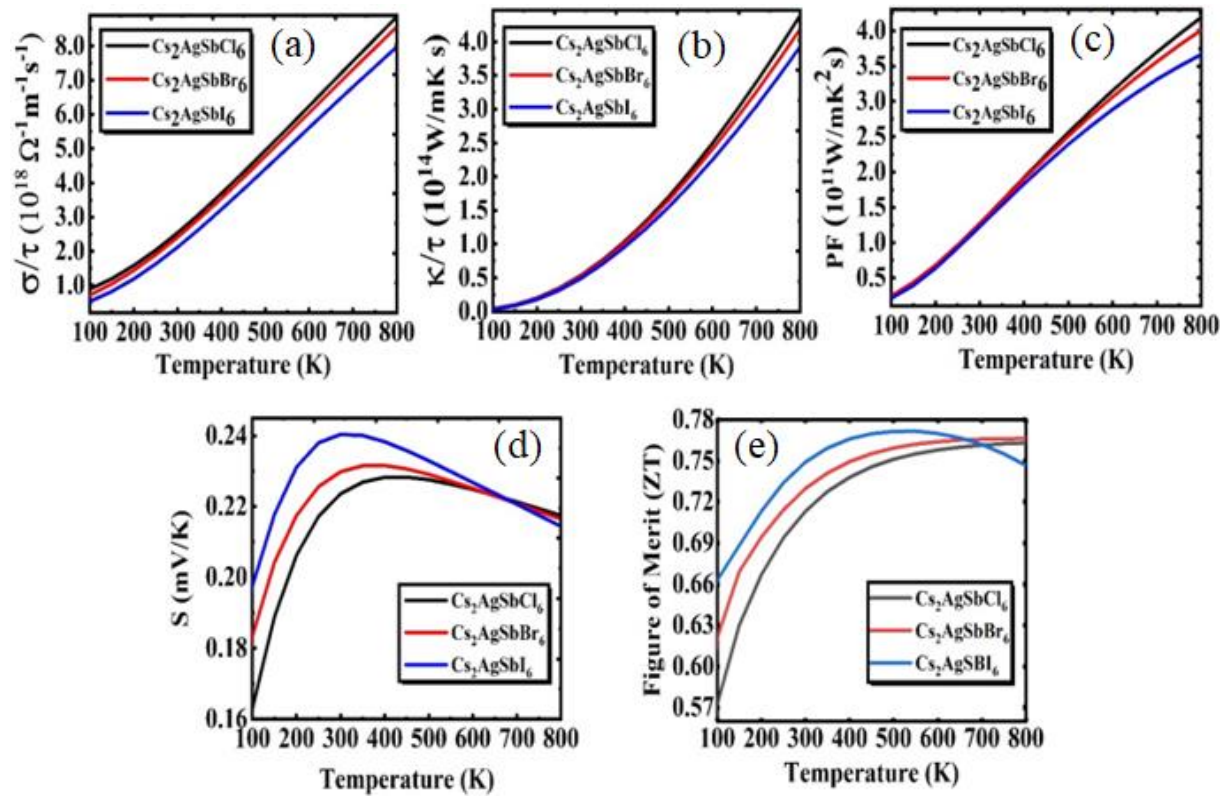

**Figure 7.** The (a) electrical conductivity (b) thermal conductivity (c) power factor (d) Seebeck coefficient and (e) figure of merit plotted as function of temperature of $Cs_2AgSbCl_6$, $Cs_2AgSbBr_6$, and $Cs_2AgSbI_6$.

## 4. Conclusions

In this work, we have investigated the structural, electronic, optical, and thermoelectric properties of double perovskites $Cs_2AgSbX_6$ (X= Cl, Br, I) using the FP-LAPW method within the DFT framework implemented in the Wien2K code. The WC-GGA approximation was used to calculate the structural parameters. All the three compounds have indirect band gap. The band gap of $Cs_2AgSbCl_6$ is 2.08 eV, which reduced to 1.37 eV and 0.64 eV for $Cs_2AgSbBr_6$ and $Cs_2AgSbI_6$, respectively, indicates their potential in single and multijunction solar cells. The broad optical absorption and optical conductivity regions for all the three compounds lie in an energy range of 6 to 12 eV, indicating their benefit in the optoelectronics devices. We found that the effective mass of electrons is smaller than that of holes resulting in a higher carrier mobility for electrons. The thermal and electrical conductivities was found to increase with temperature. The positive values of Seebeck coefficient reveal the p-type semiconductor nature of these double perovskites. The figure of merit is found to increase with temperature. The maximum values of the figure of merit are 0.77, 0.76 and 0.76 for $Cs_2AgSbI_6$, $Cs_2AgSbBr_6$ and $Cs_2AgSbCl_6$, respectively. The high electrical, small thermal conductivity, positive Seebeck coefficient, and the optimum figure of merit make these compounds suitable for thermoelectric applications.

**References**

1. Kalkan, N.;Young, E.; Celiktas, A. *Renewable and sustainable energy reviews*, **2012,** *16* (8), 6352-6383.
2. Mirza, U.K.;Ahmad, N.; Majeed, T. *Renewable and sustainable energy reviews*, **2008,** *12* (7), 1988-1996.
3. Roslan, N.;Ya'acob, M.;Radzi, M.;Hashimoto, Y.;Jamaludin, D. et al. *Renewable and sustainable energy reviews*, **2018,** *92*, 171-186.
4. Mahmood, Q.;Alhossainy, M.;Rashid, M.;Flemban, T.H.;Althib, H. et al. *Materials Science and Engineering: B*, **2021,** *266*, 115064.
5. Elsheikh, M.H.;Shnawah, D.A.;Sabri, M.F.M.;Said, S.B.M.;Hassan, M.H. et al. *Renewable and sustainable energy reviews*, **2014,** *30*, 337-355.
6. Omer, A.M. *Renewable and sustainable energy reviews*, **2008,** *12* (9), 2331-2357.
7. Ali, S.;Anwar, S.; Nasreen, S. *Forman Journal of Economic Studies*, **2017,** *13*.
8. Rodrigues, J.o.E.F.;Escanhoela Jr, C.A.;Fragoso, B.;Sombrio, G.;Ferrer, M.M. et al. *Industrial & Engineering Chemistry Research*, **2021,** *60* (51), 18918-18928.
9. Kumar, M.;Raj, A.;Kumar, A.; Anshul, A. *Optical Materials*, **2021,** *111*, 110565.
10. Jun, Z.;Ximing, R.;Molokeev, M.S.;Xiuwen, Z.; Zhiguo, X. **2018**.
11. Haque, E.; Hossain, M.A. *Journal of Alloys and Compounds*, **2018,** *748*, 63-72.
12. Ghasemi, M.;Hao, M.;Xiao, M.;Chen, P.;He, D. et al. *Nanophotonics*, **2020**.
13. Karmakar, A.;Dodd, M.S.;Agnihotri, S.;Ravera, E.; Michaelis, V.K. *Chemistry of Materials*, **2018,** *30* (22), 8280-8290.
14. Qian, F.;Hu, M.;Gong, J.;Ge, C.;Zhou, Y. et al. *The Journal of Physical Chemistry C*, **2020,** *124* (22), 11749-11753.
15. Petersen, M.;Wagner, F.;Hufnagel, L.;Scheffler, M.;Blaha, P. et al. *Computer Physics Communications*, **2000,** *126* (3), 294-309.
16. Blaha, P.;Schwarz, K.;Madsen, G.K.;Kvasnicka, D.; Luitz, J. *An augmented plane wave+ local orbitals program for calculating crystal properties*, **2001,** *60*.
17. Wu, Z.; Cohen, R.E. *Physical Review B*, **2006,** *73* (23), 235116.
18. Tran, F.; Blaha, P. *Physical review letters*, **2009,** *102* (22), 226401.
19. Madsen, G.K.; Singh, D.J. *Computer Physics Communications*, **2006,** *175* (1), 67-71.
20. Birch, F. *Physical review*, **1947,** *71* (11), 809.
21. Deng, W.;Deng, Z.-Y.;He, J.;Wang, M.;Chen, Z.-X. et al. *Applied Physics Letters*, **2017,** *111* (15), 151602.
22. García-Espejo, G.;Rodríguez-Padrón, D.;Luque, R.;Camacho, L.; de Miguel, G. *Nanoscale*, **2019,** *11* (35), 16650-16657.
23. Wei, F.;Deng, Z.;Sun, S.;Hartono, N.T.P.;Seng, H.L. et al. *Chemical Communications*, **2019,** *55* (26), 3721-3724.
24. Yu, G.;Xue, S.;Yin, R.;Wu, Q.;Gao, T. et al. *Advanced Theory and Simulations*, **2021,** *4* (8), 2100142.
25. Tran, T.T.;Panella, J.R.;Chamorro, J.R.;Morey, J.R.; McQueen, T.M. *Materials Horizons*, **2017,** *4* (4), 688-693.